\documentclass[conference]{IEEEtran}
\IEEEoverridecommandlockouts
\usepackage{cite}
\usepackage{amsmath,amssymb,amsfonts}
\usepackage{booktabs}
\usepackage{algorithmic}
\usepackage{graphicx}
\usepackage{textcomp}
\usepackage{xcolor}
\usepackage{float} % Add this line to your preamble
\usepackage{enumitem}

\def\BibTeX{{\rm B\kern-.05em{\sc i\kern-.025em b}\kern-.08em
    T\kern-.1667em\lower.7ex\hbox{E}\kern-.125emX}}
\begin{document}

\title{{MQAD}: A Large-Scale Question Answering Dataset for Training Music Large Language Models}

\author{
\IEEEauthorblockN{Zhihao Ouyang, Ju-Chiang Wang, Daiyu Zhang, Bin Chen, Shangjie Li, Quan Lin}
\IEEEauthorblockA{\textit{ByteDance} \\
oyzhouyang@gmail.com}

}

\maketitle

\begin{abstract}
Question-answering (QA) is a natural approach for humans to understand a piece of music audio. However, for machines, accessing a large-scale dataset covering diverse aspects of music is crucial, yet challenging, due to the scarcity of publicly available music data of this type. This paper introduces MQAD, a music QA dataset built on the Million Song Dataset (MSD), encompassing a rich array of musical features - including beat, chord, key, structure, instrument, and genre — across 270,000 tracks, featuring nearly 3 million diverse questions and captions. MQAD distinguishes itself by offering detailed time-varying musical information such as chords and sections, enabling exploration into the inherent structure of music within a song. To compile MQAD, our methodology leverages specialized Music Information Retrieval (MIR) models to extract higher-level musical features and Large Language Models (LLMs) to generate natural language QA pairs. Then, we leverage a multimodal LLM that integrates the LLaMA2 and Whisper architectures, along with novel subjective metrics to assess the performance of MQAD. In experiments, our model trained on MQAD demonstrates advancements over conventional music audio captioning approaches. The dataset and codes are at {https://github.com/oyzh888/MQAD}.
\end{abstract}


\section{Introduction}\label{sec:introduction}

Question-Answering (QA) systems provide an intuitive interface for interacting with and understanding music. While recent advancements in Large Language Models (LLMs), such as ChatGPT, have demonstrated ability to generate datasets in various domains \cite{bai2023qwen,chu2023qwen,liu2024visual},  applications to Music Information Retrieval (MIR) remain underexplored. As study suggests \cite{liu2024music}, training LLMs on music-specific QA data enables models to perform a wide range of tasks, reflecting what we refer to as ``emergent intelligence'' \cite{yang2023dawn,bubeck2023sparks,wei2022emergent,schaeffer2024emergent}--the ability to provide nuanced answers to complex music-related queries.

\begin{figure}[ht]
 \centerline{
 \includegraphics[width=0.9\columnwidth]{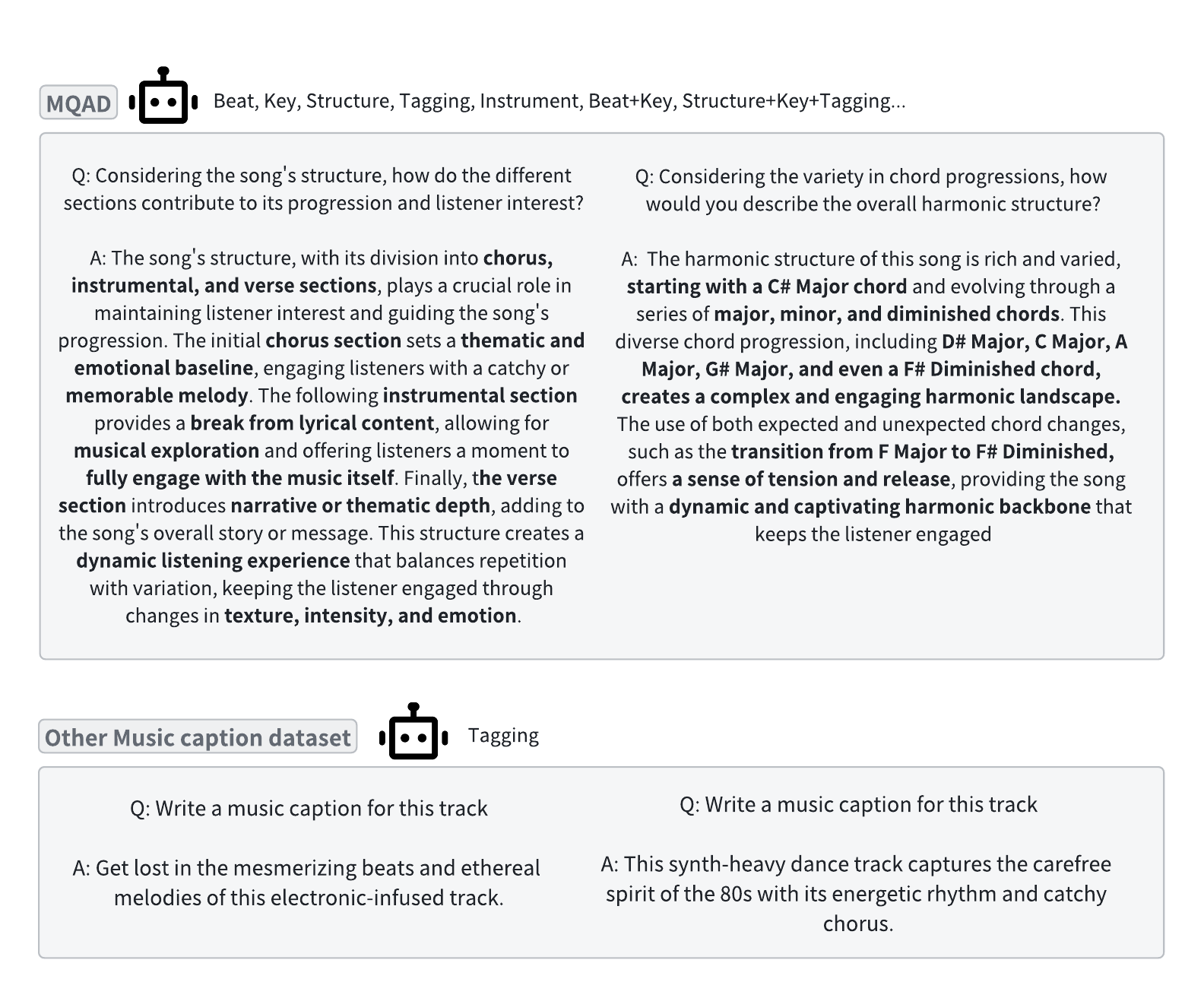}}
 \vspace{-0.2cm}
 \caption{The MQAD dataset offers diverse coverage of fine-grained MIR facets, making it ideal for training music LLMs.}
 \label{fig:mqad_dataset}
 \vspace{-0.2cm}
\end{figure}

Recent advances in multimodal LLMs (MLLMs) have shown considerable promise \cite{bai2023qwen,chu2023qwen,liu2024visual}. Foundational models for vision-language integration are detailed in \cite{zhu2023minigpt,bai2023qwen,team2023gemini,achiam2023gpt}, while \cite{chu2023qwen,liu2024music,radford2023robust,tang2023salmonn,gardner2023llark} are highlighted as seminal works within the audio domain. Despite this progress, the MIR domain lacks a robust baseline, with \cite{gardner2023llark} emerging as a foundational MLLM without offering open-source training dataset, underscoring the community's need for a comprehensive dataset to effectively train MLLMs.
While LP-MusicCaps \cite{doh2023lp} and MU-LLaMA \cite{liu2024music} have made notable contributions to the MIR community, they often fall short in handling intricate musical details such as chord progressions, song structure, and rhythmic changes. Existing datasets focus on music tagging or lack the comprehensive QA capacity needed for in-depth music analysis.

To address these limitations, we introduce MQAD, a large-scale music QA dataset based on the Million Song Dataset (MSD) \cite{bertin2011million}. MQAD includes over 270,000 tracks and nearly 3 million QA pairs and captions, covering a wide spectrum of musical features such as beats, chords, key, structural sections, and notes of multiple instruments. As illustrated in Figure \ref{fig:mqad_dataset}, the dataset enables LLMs to delve into the temporal and structural aspects of music, offering a richer understanding of its components.
Leveraging MQAD, we train a multimodal LLM that integrates the LLaMA2 \cite{touvron2023llama} and Whisper \cite{radford2023robust} architectures, yielding state-of-the-art results in music captioning and question answering tasks. Additionally, we introduce a novel evaluation metric based on GPT-4 Turbo \cite{liu2023gpteval,hsu2023gpt,hackl2023gpt,zhou2022large}, designed to simulate human judgment in assessing the quality of music QA systems.

In summary, our contributions are threefold: (1) we compile MQAD, the most extensive music QA dataset to date (with \>3 million QA pairs); (2) we develop a multimodal LLM trained on MQAD, demonstrating significant improvements in music captioning; and (3) we introduce new evaluation metrics for assessing music QA systems from multiple perspectives.

\begin{figure}
  \centerline{ 
    \includegraphics[width=0.8\columnwidth]{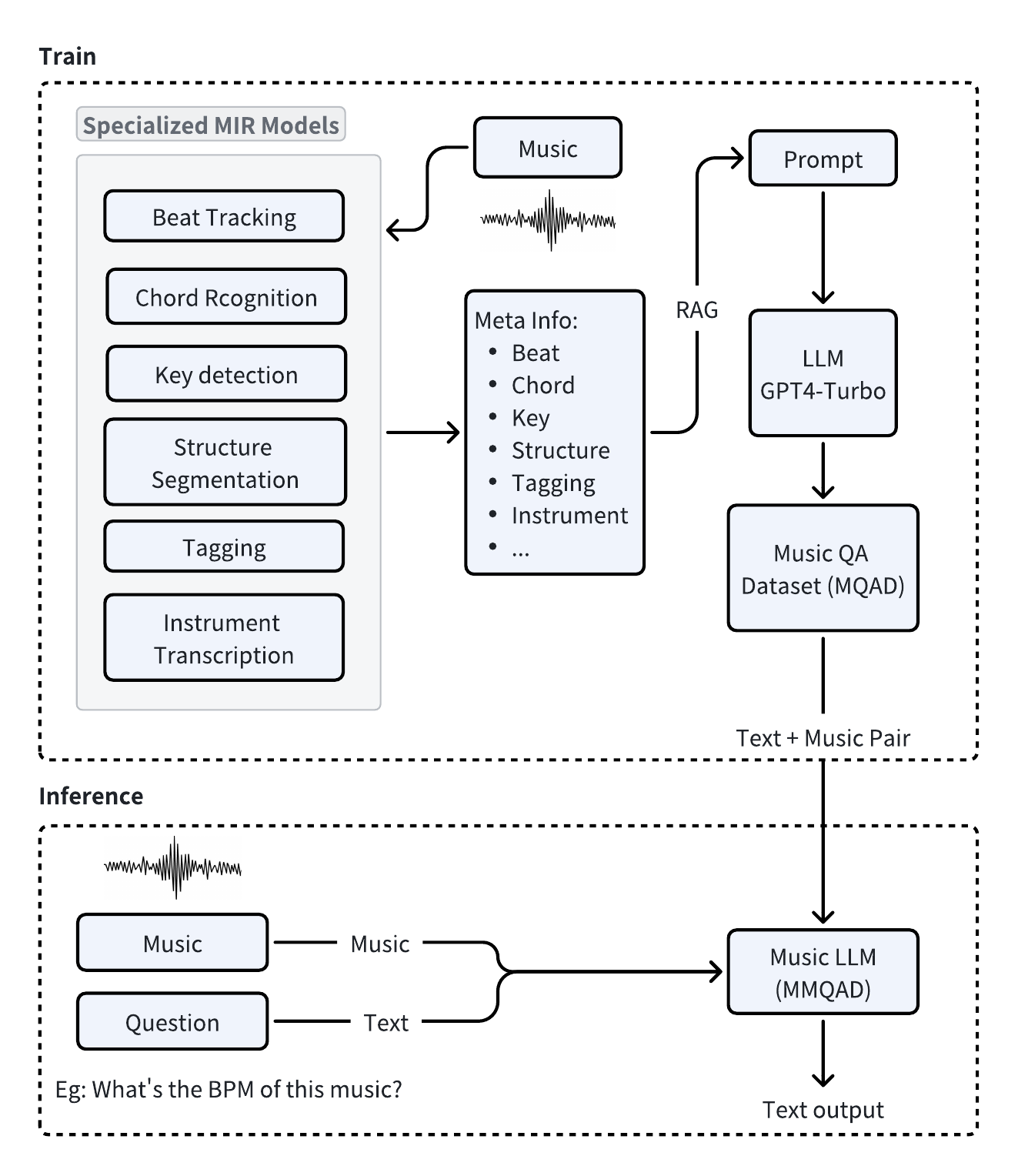} 
  } 
  \vspace{-0.2cm}
  \caption{MQAD dataset construction and MMQAD model training.} 
  \label{fig:example} 
  \vspace{-0.2cm}
\end{figure}

% \section{Related Works}
%\textbf{Related Works}~~ 
For related work, the exploration of music through the lens of QA systems has gained increasing attention alongside the development of LLMs. The MUSIC-AVQA dataset was introduced to support spatio-temporal understanding of musical content \cite{li2022learning}, offering 45K QA pairs across 33 question templates that span multiple modalities and question types. However, its primary focus is on music performance in videos, limiting its scope.
In \cite{doh2023lp}, LLMs were employed to generate captions for music, enhancing existing music tagging datasets. However, their approach is limited to general music tags, such as genre, mood, instrument, and tempo, and lacks the capability to provide nuanced temporal music information. 
Similarly, \cite{yu2023musicagent} developed a QA agent system that uses GPT-4 Turbo to classify questions, combined with an autonomous MIR model to handle specific MIR tasks. However, it did not integrate music-specific knowledge into LLM or generate a dataset that could be used by the broader community to train music-focused LLMs.
In contrast, \cite{liu2024music} introduced a model aimed at improving music captioning through a music question-answering dataset. However, this dataset is small in scale and does not address temporal aspects of music.

\section{Methodology}

\subsection{MQAD: Music QA Dataset}

% \textbf{Data source} The MQAD dataset is built upon the Million Song Dataset (MSD)\cite{bertin2011million}, leveraging its extensive collection to cover a wide range of musical genres and tags. We selected \textasciitilde 20\% of MSD according to \cite{lee2017multi}, which represents the quality samples of a total of around 270k tracks.

\noindent \textbf{Data source}~~ The MQAD dataset is built upon the Million Song Dataset (MSD) \cite{bertin2011million}, inheriting its vast array of genres and tags to ensure broad coverage. Our selection criteria is based on \cite{lee2017multi}, yielding high-quality samples of about 270k tracks in total (approximately 20\% of MSD).

\vspace{0.2cm}

\noindent \textbf{Feature Extraction}~~ 
We utilized specialized MIR models for feature extraction, including beat tracking \cite{hung2022modeling}, chord and key detection \cite{lu2021spectnt}, structure segmentation \cite{wang2022catch}, and vocal and instrument transcription \cite{lu2023multitrack}. These models are based on Transformer architecture and demonstrate state-of-the-art or comparable performance on their respective benchmarks. The extracted features are formatted as follows:
\emph{Beat}: documented in line format, each beat shows its timestamp and beat count, where a beat\_count of 1 indicates a downbeat \cite{tempobeatdownbeat:book}.
\emph{Chord}: noted with their starting and ending times and the chord name \cite{pauwels201920}.
\emph{Structure}: musical sections are listed line-by-line, each showing starting and ending times, with labels such as `intro', `verse', `chorus', `interlude', `bridge', `outro', and `silence' \cite{wang2022musfa}.
\emph{Key}: each musical section is accompanied with a key and a mode of major or minor.
\emph{Instrument Transcription}: the transcribed MIDI covers up to 12 tracks of instruments including vocals, bass, drums, guitar, and more \cite{lu2023multitrack}, formatted in JSON. Each track entry includes a list of polyphonic notes with onset and offset times and a pitch. Empty lists denote the absence of certain instruments.

These textual representations of musical events provide a rich metadata layer, enhancing the generation of QA pairs and allowing LLMs to delve into the dynamic structural compositions of music within a song.

\begin{figure}
 \centerline{
 \includegraphics[width=\columnwidth]{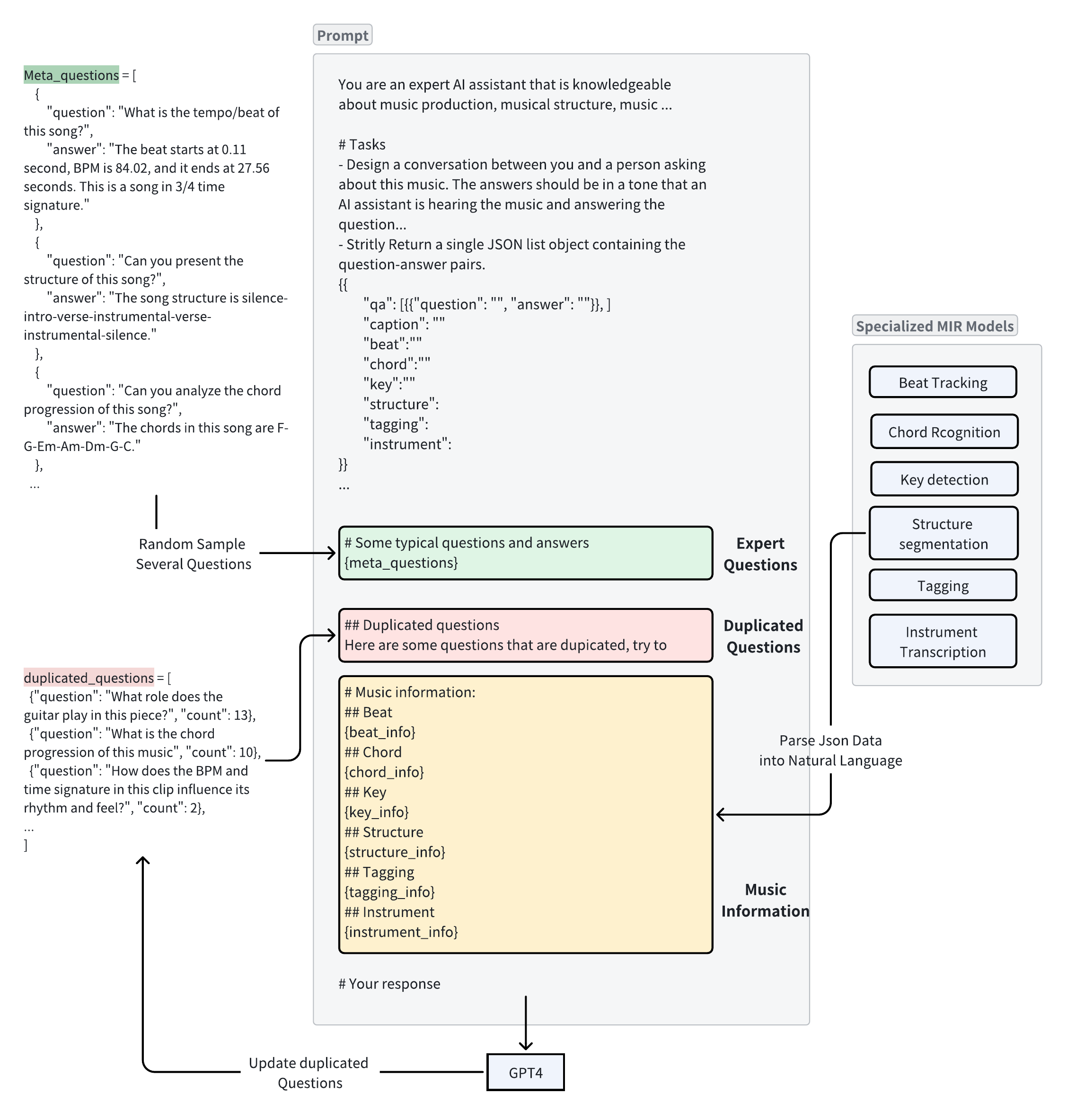}}
 \vspace{-0.2cm}
 \caption{RAG for generating QA pair and caption.}
 \label{fig:rag_system} 
 \vspace{-0.4cm}
\end{figure}

\vspace{0.2cm}

\noindent \textbf{QA Pair Generation} We used an LLM to generate text QA pairs from the extracted musical event data and the meta-information provided by MSD. To enhance the diversity and depth of the questions, we integrated a sophisticated Retrieval-Augmented Generation (RAG) system \cite{liu2023gpteval,gao2023retrieval}, as illustrated in Figure \ref{fig:rag_system}. The generation process includes:
\begin{itemize} [leftmargin=*,itemsep=0.5pt,topsep=0.5pt]
\item \emph{Backbone prompt}: Following the LLark approach  \cite{gardner2023llark}, we inform GPT with background, tasks, and system status, applying key modules to ensure high-quality, diverse content.

\item \emph{Meta Questions from Music Experts (green block)}: In collaboration with music experts, we developed meta questions covering a wide range of music aspects. Randomly incorporating these questions into prompts ensures each QA pair probes different aspects, adding specificity and depth.

\item \emph{Dynamic Prompt System (red block)}: To mitigate the tendency of LLMs like GPT-4 Turbo to generate repetitive responses, we implemented a dynamic prompting system that includes a de-duplication block to suppress questions asked previously.

\item \emph{High Diversity Parameter}: We set a high temperature (0.95) using GPT-4 Turbo to further the diversity of generated QA pairs, significantly surpassing the diversity with GPT-3.5.
\end{itemize}
These optimizations greatly reduce the question duplication rate from over 5\% to less than 0.05\%. The combined use of a RAG system, expert-derived meta questions, and GPT-4 Turbo makes MQAD one of the most comprehensive music QA datasets available.

\begin{table}[!t]
\centering
\footnotesize
\caption{Comparison of different audio-caption datasets.} 
%`C/A' is the number of captions per audio.}
\label{tab:dataset}
\vspace{-0.2cm}
\resizebox{\columnwidth}{!}{
\begin{tabular}{lrrrr}
\toprule
Dataset & \# item & Duration (h) & C/A & Avg. Token \\ \midrule
\multicolumn{5}{l}{\emph{General Audio Domain}} \\
[0.05cm]
AudioCaps \cite{kim2019audiocaps} & 51k & 144.9 & 1 & 9.0\small{±N/A} \\
LAION-Audio \cite{wu2022large} & 630k & 4325.4 & 1-2 & N/A \\
WavCaps \cite{mei2023wavcaps} & 403k & 7568.9 & 1 & 7.8\small{±N/A} \\ \midrule
\multicolumn{5}{l}{\emph{Music Domain}} \\
[0.05cm]
MusicCaps \cite{agostinelli2023musiclm} & 6k & 15.3 & 1 & 48.9\small{±17.3}\\
LP-MusicCaps-MC~\cite{doh2023lp} & 6k & 15.3 & 4 & 44.9\small{±21.3} \\
LP-MusicCaps-MTT~\cite{doh2023lp} & 22k & 180.3 & 4 & 24.8\small{±13.6} \\
LP-MusicCaps-MSD~\cite{doh2023lp} & 514k & 4283.1 & 4 & 37.3\small{±26.8} \\
\textbf{MQAD(QA)} & 804k & 11804 & 4 & 102.6\small{±23.8} \\
\textbf{MQAD-Full} & 3M & 28556 & 11 & $\sim$100 \\
\bottomrule
\end{tabular}
}
\vspace{-0.2cm}
\end{table}

\vspace{0.2cm}
\noindent \textbf{Dataset Statistics}~~ Table \ref{tab:dataset} compares various audio-caption datasets. MQAD-Full comprises 4 QA pairs and 7 captions per track in 270k MSD audio clips, totaling 3 million items—5× larger than comparable datasets and 500× larger than MusicCaps \cite{agostinelli2023musiclm}, one of the most widely used MIR datasets. An example comparison is illustrated in Figure \ref{fig:mqad_dataset}. Due to computational constraints, this work focuses on the MQAD(QA) subset, representing 36\% (4/11) of the dataset, containing only QA pairs. The remaining 64\% (7/11), mainly captions, is reserved for future use. Table \ref{tab:dataset} highlights the detailed and extensive QA pairs of MQAD, offering rich MIR information.
%As shown in Figure \ref{fig:mqad_dataset} and \ref{fig:rag_system}, our process for QA pair generation represents a leap in dataset creation methodology, particularly in the context of music. 

\subsection{MMQAD: Multimodal LLM}

We developed the MMQAD model to validate the MQAD dataset, using LLAMA2-7B \cite{touvron2023llama} as the LLM backbone and Whisper \cite{radford2023robust} as the audio encoder. Following practices in \cite{chu2023qwen,liu2024music,radford2023robust,tang2023salmonn,gardner2023llark}, input text and audio are tokenized by LLAMA2 and Whisper encoders, respectively, and processed by LLAMA2. LLAMA2’s strong performance, community support, and resource efficiency made it the ideal choice, enabling MMQAD to handle both textual and auditory inputs for diverse MIR tasks.

The training of MMQAD was conducted on the MQAD dataset, utilizing a subset (i.e., `qa' key) of the question set specifically tailored for our QA use cases. For all experiments, the input to the encoder is an audio clip of up to 30 seconds at a 16kHz sampling rate, converted to a log-scaled mel spectrogram with 80 mel bins, a 25 ms Hann window, a 10 ms stride, and a 10 ms hop size. All models were trained using the AdamW optimizer with a learning rate of 1e-4. We employed a cosine learning rate decay to zero after a warm-up period of 1000 steps. Two scenarios of training processes are considered: pre-training and fine-tuning.
For pre-training, we used a batch size of 256, and the models were trained for 32,768 steps. For fine-tuning and transfer learning, we used a batch size of 64 and trained for 10 epochs. Beam search with 5 beams was employed for the inference of all models. We trained the self-attention layers using LoRA \cite{hu2021lora} and froze the Whisper encoder to conserve computational resources. For fine-tuning, we slightly decreased the learning rate.

\subsection{Subjective and Objective Metrics}
For evaluation, we focus on objective metrics as suggested in \cite{doh2023lp}, which include BLEU-1 to BLEU-4 (B1--4), METEOR (M), ROUGE-L (R-L), and BERT Score (BERT). We also propose a (pseudo-) subjective metric that leverages LLM as a judger. Research in NLP has shown a high correlation between human judgment and assessments made by LLMs \cite{liu2023gpteval,hsu2023gpt,hackl2023gpt,zhou2022large}, supporting the use of this metric to evaluate the overall quality of MMQAD outputs from a comprehensive perspective.
Specifically, we employed GPT4-Turbo to compare predicted answers with the ground truth across eight distinct musical dimensions: Average Accuracy, Average Keywords Matching, and various Average Music Metrics including beat tracking, structural segmentation, chord and key detection, instrumentation, genre, and cultural appropriateness. 
%This approach not only assesses the overall quality of the QA but also delves into both the broad and nuanced aspects of musical understanding.

\begin{table}[t]
\centering
\footnotesize
\caption{Music captioning results on the MusicCaps test set.}
\label{tab:music_captioning_results}
\vspace{-0.2cm}
\begin{tabular}{p{2.51cm}p{0.4cm}p{0.4cm}p{0.4cm}p{0.5cm}p{0.5cm}p{1.3cm}}
\hline
\textbf{Model} & \multicolumn{5}{c}{\textbf{Supervised Metrics(\%)}} & \multicolumn{1}{c}{\textbf{Length}} \\
 & B1 & B2 & M & R-L & BERT & Avg.Tokens \\
\hline
\multicolumn{7}{l}{\textbf{Baseline}} \\
[0.05cm]
Supervised Model & 28.51 & 13.76 & 20.62 & 19.22 & 87.05 & 46.7$\pm$16.5  \\
\midrule
\multicolumn{7}{l}{\textbf{Pre-training (Zero-shot Captioning)}} \\
[0.05cm]
Tag Concat [2, 13] & 4.33 & 0.84 & 3.10 & 2.01 & 79.30 & 23.8$\pm$12.1 \\
Template [14] & 7.22 & 1.58 & 5.28 & 6.81 & 81.69 & 25.8$\pm$12.4 \\
K2C-Aug [22] & 7.67 & 2.10 & 7.94 & 11.37 & 82.99 & 19.9$\pm$7.6 \\
LP-MusicCaps\cite{doh2023lp} & 19.77 & 6.70 & 12.88 & 13.03 & 84.51 & 45.3$\pm$28.0 \\
MMQAD-B (Ours) & 13.16 & 4.18 & 11.24 & 13.91 & 85.18 & 24.9$\pm$5.4 \\
MMQAD-C (Ours) & \textbf{21.55} & \textbf{7.16} & \textbf{19.41} & \textbf{15.94} & \textbf{85.41} & \textbf{75.3$\pm$12.1} \\
MMQAD-D (Ours) & 10.65 & 3.17 & 9.52 & 12.60 & 85.19 & 28.0$\pm$10.5 \\
\midrule
\multicolumn{7}{l}{\textbf{Fine-tuning (Transfer Learning)}} \\
[0.05cm]
Tag Concat [2, 13] & 28.65 & 14.68 & 21.88 & 21.31 & 87.67 & 41.8$\pm$14.3 \\
Template [14] & 28.41 & 14.49 & 21.88 & 21.25 & 87.72 & 41.1$\pm$13.2 \\
K2C-Aug [22] & 29.50 & 14.99 & 21.97 & 20.92 & 87.50 & 44.1$\pm$15.0 \\
LP-MusicCaps\cite{doh2023lp} & 29.09 & 14.87 & \textbf{22.39} & 21.49 & 87.78 & 42.5$\pm$14.3 \\
MMQAD-C+F (Ours) & \textbf{30.30} & \textbf{15.49} & 22.25 & \textbf{21.56} & \textbf{87.78} & \textbf{45.12$\pm$13.71} \\
% MMQAD-B+F (Ours) & \textbf{30.31} & \textbf{15.50} & 8.81 & 5.64 & 22.25 & \textbf{21.58} & \textbf{87.78} & \textbf{45.12$\pm$13.71} \\
\hline
\end{tabular}
\vspace{-0.2cm}
\end{table}

\begin{table*}[h]
\footnotesize
\centering
\caption{Music QA results on the MQAD test set.}
\vspace{-0.2cm}
\label{tab:music_captioning_results_mqad}
\begin{tabular}{lccccccccc}
\hline
\textbf{Model} & \multicolumn{7}{c}{\textbf{Supervised Metrics(\%)}} & \multicolumn{1}{c}{\textbf{Length}} \\
 & \textbf{B1} & \textbf{B2} & \textbf{B3} & \textbf{B4} & \textbf{M} & \textbf{R-L} & \textbf{BERT-S} & \textbf{Avg.Token} \\
\hline
LP-MusicCaps\cite{doh2023lp} & 15.24 & 4.74 & 1.17 & 0.28 & 12.14 & 13.72 & 84.48 & \(50.12 \pm 14.06\) \\
MMQAD-B (Ours) & \textbf{51.86} & \textbf{35.01} & \textbf{25.59} & \textbf{19.78} & \textbf{40.23} & \textbf{36.13} & \textbf{91.32} & \(\mathbf{75.14 \pm 13.95}\) \\
MMQAD-C (Ours) & 51.47 & 34.65 & 25.35 & 19.58 & 40.04 & 35.77 & 91.31 & \(74.85 \pm 14.06\) \\
MMQAD-D (Ours) & 7.33 & 2.18 & 0.72 & 0.25 & 9.78 & 13.39 & 85.49 & \(32.93 \pm 18.31\) \\
% MMQAD-B+F (Ours) & 10.68 & 3.85 & 1.17 & 0.43 & 11.64 & 15.53 & 84.68 & \(39.47 \pm 11.08\) \\
% 这个结果挺有意思的，但是我想可以放到supplement materials里，可以说明数据集之间确实存在冲突，finetune musiccap-mc-train是对QA任务不好的。
\hline
\end{tabular}
\vspace{-0.2cm}
\end{table*}

\begin{table}[t]
\centering
\footnotesize
\caption{Subjective music QA results of models on MQAD test set.}
\label{tab:mqad_gpt_eval}
\vspace{-0.2cm}
\begin{tabular}{lccc}
\hline
\textbf{Metric(\%)} & -B & -C & -D \\
\hline
Accuracy & \textbf{87} & 86 & 26 \\
Keywords Match & \textbf{89} & 87 & 28 \\
Beat Tracking & \textbf{85} & \textbf{85} & 24 \\
Structural Segment & \textbf{87} & \textbf{87} & 22 \\
Key Detection & \textbf{82} & 80 & 21 \\
Genre and Mood & \textbf{86} & 85 & 32 \\
Instrument Presence & \textbf{85} & \textbf{85} & 28 \\
Cultural Appropriateness & \textbf{77} & 76 & 22 \\
\hline
\end{tabular}
\end{table}

\section{Experiments}
We compared MMQAD and MusicCaps-LP models using the test sets of LP-MusicCaps-MC and MQAD.

\subsection{Datasets and Models}

In this work, three datasets are studied: LP-MusicCaps-MSD, LP-MusicCaps-MC, and our MQAD.
\emph{LP-MusicCaps-MSD} consists of approximately 445K training examples. This dataset serves as a pre-training resource for music captioning tasks.
\emph{LP-MusicCaps-MC} is more concise dataset with about 2.6K training examples and 2.8K test examples. This is used to assess the model's performance in both supervised and zero-shot settings.
\emph{MQAD} is our primary dataset, featuring approximately 804K training QA pairs derived from 213K songs. The validation set contains about 46K QA pairs from 12K songs, and the test set includes 110K QA pairs across 28K songs, with a subset of 500 high-quality questions, providing a wide spectrum for comprehensive evaluation.

Since QA tasks include captioning, we adapted the LP-MusicCaps-MC test set into a QA format for fair comparison. This was achieved by appending the prompt \emph{“write a music caption for this track”} to allow the MMQAD model to process it as a music captioning task. In contrast, MusicCaps-LP, being a native music captioning model designed exclusively for audio input, does not require any textual prompts.

%\noindent \textbf{Model Pre-training}
For pre-training, three models are defined as follows: 
\emph{MMQAD-B} is pre-trained on the combination of LP-MusicCaps-MSD training set and MQAD training set.
\emph{MMQAD-C} is pre-trained exclusively on the MQAD training set.
\emph{MMQAD-D} is pre-trained on the LP-MusicCaps-MSD training set alone.
For fine-tuning, we define \emph{MMQAD-C+F}, where we further fine-tuned \emph{MMQAD-C} on LP-MusicCaps-MC training set, to verify improvement in music captioning under supervised conditions.

%\subsection{Result and Comparison}

\subsection{Result for Music Captioning Task}

Table \ref{tab:music_captioning_results} shows the music captioning performance on the LP-MusicCaps-MC test set. In \textbf{pre-training} setting, \emph{MMQAD-C} outperforms other models, particularly in METEOR and ROUGE-L scores, underscoring the advantages of utilizing the MQAD dataset.
In \textbf{fine-tuning} setting, the models enhanced through fine-tuning generally show improved performance, with \emph{MMQAD-C+F} particularly excelling. This highlights the MQAD effectiveness in refining music captioning tasks.

Comparing between \emph{MMQAD-C} and \emph{MMQAD-D}, we observe differences between MQAD and LP-MusicCaps-MSD, suggesting that combining datasets, as in MMQAD-B, may not yield significant performance improvements. This highlights that MMQAD, leveraging a large-scale LLM like LLAMA2-7B, requires diverse QA data for effective training. The fixed-question approach in LP-MusicCaps-MSD may limit its performance. For details on the `Supervised Model’ in Table \ref{tab:music_captioning_results}, see \cite{doh2023lp}.
%Furthermore, we observe MMQAD's performance appears to be inferior to other models in BLEU-3 and BLEU-4. 
Furthermore, we observe the tendency for MMQAD to generate more detailed responses, averaging 75.3 tokens per output. This verbosity would possibly affect the results of longer n-grams (e.g., BLEU-3 and BLEU-4), where the lengthy generated text results in lower scores despite the correctness of the content. In contrast, shorter n-grams like BLEU-1 are less impacted by such verbosity.
%\noindent \textbf {QA Performance}

\subsection{Results for the Music QA Task}
Evaluating the music question-answering task poses significant challenges, necessitating both subjective and objective approaches. Given the limited scope of the LP-MusicCaps-MC, which contains fewer than 6,000 entries, it does not sufficiently challenge a model's comprehensive QA capabilities. Therefore, we compiled the MQAD test set, which includes 100K samples featuring detailed MIR questions such as chord progression and music structure.

For \textbf{objective evaluation}, Table \ref{tab:music_captioning_results_mqad} presents the QA performance comparison on the MQAD test set. Comparing between LP-MusicCaps and MMQAD shows that MMQAD-B achieves the best results across all metrics, slightly surpassing MMQAD-C. However, MMQAD-D underperforms significantly due to its training data being limited to music captions, which do not adequately test its broader QA capabilities.

For \textbf{subjective evaluation}, considering the costs associated with using GPT-4, we selected 500 representative cases from our MQAD test set that span a broad spectrum of MIR questions. As indicated in Table \ref{tab:mqad_gpt_eval}, MMQAD-B stands out as the top performer, which is in line with the findings in Table \ref{tab:music_captioning_results_mqad}. Overall, our models consistently exhibit superior performance across various MIR dimensions, demonstrating the efficacy of our testing suite in evaluating QA quality across a diverse range of metrics. 

\section{Conclusion}
We have presented the MQAD dataset and the MMQAD model, which establish a new paradigm for research in MIR by leveraging the power of large-scale, diverse QA datasets and multimodal LLMs. 
%By offering a detailed exploration of music through the lens of question-answering systems, we have demonstrated the potential for significant advancements in how machines understand and interact with music. 
For future work, our dataset could enhance text-to-music generation by integrating with MMQAD, enabling more nuanced user queries for temporal control, which may improve the quality and precision of generated outputs.
%As we look to the future, the paths for continued research and development are abundant, promising even more profound insights and capabilities in the realm of music AI.

\vfill\pagebreak

\label{sec:refs}

% \bibliographystyle{IEEEbib}
% \bibliography{strings,refs}

\bibliographystyle{IEEEbib}
\fontsize{9}{10}\selectfont % This reduces the font size of the bibliography to 9pt
\bibliography{strings,refs}

\end{document}